\documentclass[prl,superscriptaddress,unsortedaddress,twocolumn,showpacs,
floatfix,preprintnumbers]{revtex4} 

\usepackage{graphicx}
\usepackage{dcolumn}
\usepackage{bm}

\newcommand{\mee}{$M_{e^+e^-}$}
\newcommand{\ketee}{$K_L\rightarrow \pi^{\pm}e^{\mp} \nu e^+e^-$}
\newcommand{\keteem}{$K_{e3ee}$}
\newcommand{\kesan}{$K_{e3}$}
\newcommand{\pmzd}{$K_L\rightarrow \pi^+\pi^-\pi^0_D$}
\newcommand{\pmz}{$K_L\rightarrow \pi^+\pi^-\pi^0$}
\newcommand{\pmzgg}{$K_L\rightarrow \pi^+\pi^-\pi^0_{\gamma \gamma}$}
\newcommand{\kefd}{$K_L \rightarrow \pi^{\pm}e^{\mp}\nu\pi^0_D$}

\newcommand{\pmzfe}{$K_L\rightarrow \pi^+\pi^-\pi^0_{4e}$}
\newcommand{\pizeeg}{$\pi^0\rightarrow e^+e^- \gamma$}
\newcommand{\kpi}{$K$-$\pi$}

\newcommand{\eop}{$E/p$}
\newcommand{\eepair}{$e^+e^-$ pair}
\newcommand{\ketg}{\kesan$_\gamma$}

\newcommand{\nlopf}{NLO($p^4$)}
\newcommand{\eket}{$e^{\pm}_{ke3}$}

\newcommand{\pieee}{$\pi^{\pm} e^{\mp} e^+ e^- $}

\newcommand{\cascade}{$\Xi \rightarrow \Lambda (\rightarrow p \pi^-) \pi_D^0$}
\newcommand{\ppz}{$k_{+-0}$}

\begin{document}


\title{First observation of  
$\bm{K_L\rightarrow \pi^{\pm}e^{\mp}\nu e^+e^-}$}


\newcommand{\UAz}{University of Arizona, Tucson, Arizona 85721}
\newcommand{\UCLA}{University of California at Los Angeles, Los Angeles,
                    California 90095} 
\newcommand{\Campinas}{Universidade Estadual de Campinas, Campinas, 
                       Brazil 13083-970}
\newcommand{\EFI}{The Enrico Fermi Institute, The University of Chicago, 
                  Chicago, Illinois 60637}
\newcommand{\UB}{University of Colorado, Boulder, Colorado 80309}
\newcommand{\ELM}{Elmhurst College, Elmhurst, Illinois 60126}
\newcommand{\FNAL}{Fermi National Accelerator Laboratory, 
                   Batavia, Illinois 60510}
\newcommand{\Osaka}{Osaka University, Toyonaka, Osaka 560-0043 Japan} 
\newcommand{\Rice}{Rice University, Houston, Texas 77005}
\newcommand{\SaoPaolo}{Universidade de Sao Paolo, Sao Paulo, Brazil 05315-970}
\newcommand{\UVa}{The Department of Physics and Institute of Nuclear and 
                  Particle Physics, University of Virginia, 
                  Charlottesville, Virginia 22901}
\newcommand{\UW}{University of Wisconsin, Madison, Wisconsin 53706}

\affiliation{\UAz}
\affiliation{\UCLA}
\affiliation{\Campinas}
\affiliation{\EFI}
\affiliation{\UB}
\affiliation{\ELM}
\affiliation{\FNAL}
\affiliation{\Osaka}
\affiliation{\Rice}
\affiliation{\SaoPaolo}
\affiliation{\UVa}
\affiliation{\UW}

\author{E.~Abouzaid}	  \affiliation{\EFI}
\author{M.~Arenton}       \affiliation{\UVa}
\author{A.R.~Barker}      \altaffiliation[Deceased.]{ } \affiliation{\UB}
\author{L.~Bellantoni}    \affiliation{\FNAL}
\author{E.~Blucher}       \affiliation{\EFI}
\author{G.J.~Bock}        \affiliation{\FNAL}
\author{E.~Cheu}          \affiliation{\UAz}
\author{R.~Coleman}       \affiliation{\FNAL}
\author{M.D.~Corcoran}    \affiliation{\Rice}
\author{B.~Cox}           \affiliation{\UVa}
\author{A.R.~Erwin}       \affiliation{\UW}
\author{C.O.~Escobar}     \affiliation{\Campinas}  
\author{A.~Glazov}        \affiliation{\EFI}
\author{A.~Golossanov}    \affiliation{\UVa} 
\author{R.A.~Gomes}       \affiliation{\Campinas}
\author{P. Gouffon}       \affiliation{\SaoPaolo}
\author{Y.B.~Hsiung}      \affiliation{\FNAL}
\author{D.A.~Jensen}      \affiliation{\FNAL}
\author{R.~Kessler}       \affiliation{\EFI}
\author{K.~Kotera}	  \affiliation{\Osaka}
\author{A.~Ledovskoy}     \affiliation{\UVa}
\author{P.L.~McBride}     \affiliation{\FNAL}

\author{E.~Monnier}
   \altaffiliation[Permanent address ]{C.P.P. Marseille/C.N.R.S., France}
   \affiliation{\EFI}  

\author{K.S.~Nelson}     \affiliation{\UVa}  
\author{H.~Nguyen}       \affiliation{\FNAL}
\author{R.~Niclasen}     \affiliation{\UB}
\author{D.G.~Phillips~II} \affiliation{\UVa}
\author{H.~Ping}         \affiliation{\UW}  
\author{E.J.~Ramberg}    \affiliation{\FNAL}
\author{R.E.~Ray}        \affiliation{\FNAL}
\author{M.~Ronquest}     \affiliation{\UVa}
\author{E.~Santos}       \affiliation{\SaoPaolo}
\author{W.~Slater}       \affiliation{\UCLA}
\author{D.~Smith}        \affiliation{\UVa}
\author{N.~Solomey}      \affiliation{\EFI}
\author{E.C.~Swallow}    \affiliation{\EFI}\affiliation{\ELM}
\author{P.A.~Toale}      \affiliation{\UB}
\author{R.~Tschirhart}   \affiliation{\FNAL}
\author{C.~Velissaris}   \affiliation{\UW}  
\author{Y.W.~Wah}        \affiliation{\EFI}
\author{J.~Wang}         \affiliation{\UAz}
\author{H.B.~White}      \affiliation{\FNAL}
\author{J.~Whitmore}     \affiliation{\FNAL}
\author{M.~J.~Wilking}      \affiliation{\UB}
\author{B.~Winstein}     \affiliation{\EFI}
\author{R.~Winston}      \affiliation{\EFI}
\author{E.T.~Worcester}  \affiliation{\EFI}
\author{M.~Worcester}    \affiliation{\EFI}
\author{T.~Yamanaka}     \affiliation{\Osaka}
\author{E.~D.~Zimmerman} \affiliation{\UB}
\author{R.F.~Zukanovich} \affiliation{\SaoPaolo}
\collaboration{The KTeV Collaboration}

\date{May 21, 2007}

\begin{abstract}
This letter is the first report of the  $K_L\rightarrow \pi^{\pm}e^{\mp}\nu e^+e^-$ decay.
Based on $19208\pm144$ events, we determine the branching fraction,
$B(K_L\rightarrow \pi^{\pm}e^{\mp}\nu e^+e^-; M_{e^+e^-} > 5\ $MeV$/c^2,E^*_{e^+e^-} > 30\ $MeV)$=(1.285\pm0.041)\times10^{-5}$, and $\Gamma(K_{e3ee}; 
	M_{e^+e^-} > 5 {\rm MeV}/c^2) /
	\Gamma(K_{e3})
	=  [4.57\pm0.04(stat)\pm0.14(syst)]\times10^{-5}$.
This ratio agrees with a theoretical prediction based on chiral perturbation theory
(ChPT) calculated to $\mathcal{O}(p^4)$.
The measured kinematical distributions agree with those predicted by 
just ChPT $\mathcal{O}(p^4)$, but show significant disagreement with ones predicted 
by leading order ChPT.
\end{abstract}

\pacs{13.20.-v, 13.20.Eb, 13.25.Es, 14.40.Aq, 12.39.Fe}

\maketitle

\begin{figure}[htb]									
\begin{center}\includegraphics[width=8.5cm]{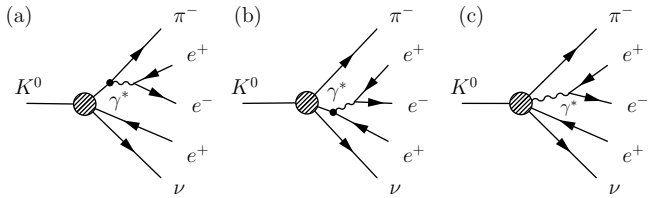}		
\caption{\label{fig:diagrams}
{Processes contributing to the \ketee\ decay: the virtual photon comes (a)
from the charged pion, (b) from the \kesan\ electron, and (c) from the kaon decay vertex.
The process (c) includes the structure dependent (SD) amplitudes.  The exchange diagrams
have been omitted for clarity.
%
}}
\end{center} 
\end{figure}

The semileptonic mode, $K_L\rightarrow \pi^{\pm}e^{\mp}\nu\ $($K_{e3})$
and its radiative mode, $K_{e3\gamma}$, has been extensively studied
\cite{Fearing:1970zz, Alavi-Harati:2001wd, Alexopoulos:2004up}.
In this letter,
we introduce the semileptonic kaon decay mode
$K_L\rightarrow \pi^{\pm}e^{\mp}\nu e^+e^-\ $(\keteem).
We present the first measurement of its branching fraction, and the ratio of its 
decay width to that of the \kesan\ decay.  As shown in Fig.~\ref{fig:diagrams},
the \ketee\ decay is dominated by \kesan\ with a virtual photon, $\gamma^*$, 
that converts internally into a real \eepair.
%
The amplitude of \ketg\ consists of two parts.
One part is inner bremsstrahlung (IB) from the pion or the electron,
which is illustrated by Fig.~\ref{fig:diagrams}-(a) and (b). 
The other part is the photon radiated from an intermediate hadronic state of 
the \kpi\ current, namely the structure dependent (SD) amplitude (ie. direct emission) 
\cite{Fearing:1970zz, Alavi-Harati:2001wd, Gasser:2004ds}, as illustrated by 
Fig.~\ref{fig:diagrams}-(c).
\par
The model describing the \kpi\ current is important for  
both studying \kesan\ decays themselves and understanding low energy QCD.
A powerful way to express the \kpi\ current is via Chiral Perturbation Theory (ChPT) 
\cite{daphne_1994}. 
ChPT has been developed based on the chiral symmetry of QCD, and 
it can be applied to all \kesan\ modes, including \keteem.
In this letter, we compare our measurements against ChPT calculated 
to next-to-leading order, expanded 
to the fourth power of the momentum of chiral field $p$ [\nlopf].


We search for \keteem\ decays in data collected by the
KTeV E799-II fixed target experiment, which ran at the Fermi National Accelerator Laboratory.
Two parallel $K_L$ beams were produced by 800 GeV/c protons from the Tevatron striking a BeO target.
Following the target were beam line elements to sweep away charged particles, 
to absorb photons, and to allow for short-lived hadrons to decay away.
The region from 95 m to 159 m downstream of the target was in vacuum, and defines the 
fiducial volume for $K_L$ decays.
Following a thin vacuum window at the end of the decay region was a drift chamber spectrometer.
The spectrometer had two pairs of drift chambers separated by an analysis magnet 
providing a transverse momentum kick of 0.2 GeV/c.
The momentum resolution of the spectrometer was measured to be 
 $\sigma_p/p = 0.016\% \times p \oplus 0.38\%$, where $p$ is the momentum of a charged particle
 in GeV/c.  A set of transition radiation detectors (TRD) 
downstream of the spectrometer was used to distinguish pions and electrons.
Farther downstream, there was a pure cesium iodide (CsI) electromagnetic calorimeter,
where the energy resolution for photons and electrons was
$\sigma_E/E = 2\% / \sqrt{E} \oplus 0.45\%$, with $E$ in GeV.
Immediately upstream of the CsI calorimeter were scintillator hodoscopes, which served as the 
charged particle trigger.  Behind the CsI was a 24 interaction-length steel filter and 
a set of scintillator hodoscopes to identify muons.  
Lead-scintillator counters were positioned around the vacuum decay region, the spectrometer and 
the calorimeter, to reject events with particles escaping these detectors.
We analyzed data acquired during the 1997 run.
A detailed description for this experiment and analysis can be found in Ref.~\cite
{Kotera:2006vn, Alavi-Harati:2002ye}.


The event reconstruction begins with the identification of four charged tracks 
coming from a vertex in the decay region.
The charged tracks are identified as \pieee\ using \eop, 
the energy reconstructed in the CsI calorimeter divided by the momentum measured in the 
spectrometer. 
Pions tracks are required to have $E/p$ less than 0.9, which selects 99.2\% of all pions. 
Electron tracks are required to have $E/p$ between 0.93 and 1.15, and be 
tagged by the TRD system.  The \eop\ and TRD requirements select 95.0\% of all electrons,
while rejecting 99.95\% of all pions.
Since the \keteem\ decay has three electrons, there are two candidates for the \eepair.
Although each event must include both amplitudes 
in which one of \eepair s  is from the virtual photon,
in this letter, we define the pair which has smaller invariant mass as the ``\eepair", and call 
the remaining electron as ``\eket".
The amplitude in which the smaller invariant mass \eepair\ emerges from 
the virtual photon is greater in contribution than the other amplitudes.
Because the neutrino is not observed, there is a two-fold ambiguity for the parent kaon energy.
The higher kaon energy solution was required to be less than 200 GeV.


Monte Carlo (MC) simulations are used to understand the acceptance of the signal mode,
background modes, and the normalization mode \pmzd\, where $\pi_D^0$ denotes the \pizeeg\ decay.
To simulate \keteem, the matrix element was computed using ChPT[\nlopf]
as described by Tsuji $et\ al.$\cite{Tsuji}.
Bremsstrahlung photons from four charged particles in \keteem\ are
simulated using the {\sc photos} program,
which includes the interference terms \cite{Barberio:1993qi, Was:2004dg}.
The number of background MC events are generated according to their branching fractions
and $K_L$ flux estimated using the normalization mode. 


\begin{figure}[htb]									
\begin{center}\includegraphics[width=7cm]{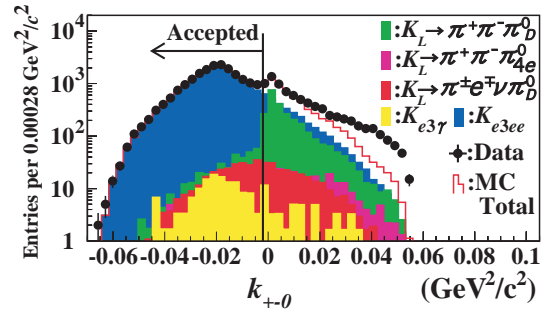}	
\caption{\label{fig:pp0}
{ The \ppz\ distributions for data and MC after all analysis requirements except for `\ppz'.
The open histogram is the sum of all MC simulated modes (MC Total).  
The vertical line and arrow show the accepted region for the signal candidates (\ppz $
< -0.002$ GeV$^2$/c$^2$).
}}
\end{center} 
\end{figure} 

\begin{table*}
\caption{\label{tbl:background} The estimated backgrounds (BG) for the \keteem\ analysis and the
dependence on analysis cuts. The third column show the BG/Signal reduction 
factor stemming from the combined particle ID, fiducial, and veto cuts.  The fourth 
column shows the same effect for the kinematic cut. }
\begin{ruledtabular}
\begin{tabular}{lddcc} 
Background & \mbox{BG/Signal} &\hspace{-10mm} \mbox{BG/Signal}& 
\multicolumn{2}{c}{BG/Signal Reduction Factor} \\ \cline{4-5}   
decay mode & \mbox{before cuts} &\hspace{-10mm}  \mbox{after all cuts}  
& Particle ID+Fiducial+Veto Cuts & Kinematic cut \\  
\hline
\pmzd & 117 & \hspace{-10mm}0.0182 & 403 & 16 \\ 
\kefd & 0.048 & \hspace{-10mm}0.0167 & 1.23 & 2.4 \\ 
\pmzfe & 0.307 & \hspace{-10mm}0.0096 & 4.48 & 7.2 \\  
$K_L \to \pi^{\pm} e^{\mp} \nu \gamma$, $\gamma$ conversion & 15.0
\footnote{The minimum simulated photon energy was 1 keV.}
 & \hspace{-10mm}0.0081& 20.4 & 91 \\  
\end{tabular}
\end{ruledtabular}
\end{table*}

The backgrounds are reduced with a combination of particle identification, kinematical cuts,
and vetos as indicated in Table \ref{tbl:background}.
The major background for the \keteem\ mode is due to \pmzd.
MC studies show that 42\% of the background arising from the \pmzd\ decays are caused 
by one of the pions misidentified as an electron. 
The rest are due to an external photon conversion in the detector material and 
a missing pion and electron. 
Missing pions are due to track hits corrupted by
hadronic interactions in the detector material, while the analysis magnet causes 
low momentum tracks to escape the detector. 
We suppress the \pmzd\ background using the kinematical variable, 
\begin{equation}
k_{+-0} = \frac{(M_K^2-M_{\pi e_{ke3}}^2-M_{\pi^0}^2)^2-4M_{\pi e_{ke3}}^2M_{\pi^0}^2-4M^2_Kp_t^2}
			{4(M_{\pi e_{ke3}}^2 + p_t^2)},
\end{equation}
where $M_K$ and $M_{\pi^0}$ are the kaon and $\pi^0$ masses, respectively. 
$M_{\pi e_{ke3}}$ is the invariant mass of the $\pi^{\pm}$ and $e_{ke3}^{\mp}$,
while the charged pion mass is assigned to the $e_{ke3}^{\mp}$.
$p_t$ is the transverse momentum of the $\pi^{\pm} e_{ke3}^{\mp}$ system.
For \pmz\ decays, \ppz\ is the squared longitudinal momentum of the $\pi^0$ in the 
frame in which the momentum of $\pi^+\pi^-$ system is transverse to the $K_L$ direction,
so that \ppz\ will be positive definite (Fig.~\ref{fig:pp0}).
On the other hand, for \keteem\ events, \ppz\ tends to have an unphysical value (\ppz$<0)$.
Therefore, we require \ppz\ $< -0.002$ GeV$^2/c^2$ for the signal events.
The radiative \kesan\ decay with an external photon conversion 
are rejected by requiring $M_{e^+e^-} > 5$ MeV/c$^2$.
We also considered and simulated backgrounds due to \kefd\ and \pmzfe.  
These backgrounds are small mainly because of the small branching fractions.
\par
In addition to the backgrounds in Table \ref{tbl:background}, 
we also considered and simulated two coincident $K_L \to \pi^{\pm} e^{\mp} \nu$ decays, 
and the  \cascade\ decay;  both of these backgrounds are negligible.
After all cuts, we are left with a sample of 20225 events.  
The estimated total number of background events
after all cuts is $1017.1\pm24.7$, representing ($5.03\pm0.12$)\%
of the signal sample.


The normalization mode (\pmzd) were collected with the same conditions as the signal mode.
They are analyzed using identical cuts, 
except that the \ppz\ requirement is reversed to be \ppz$ > -0.002$ GeV$^2$/c$^2$.
We ignore the photon in the decay in order to make the analysis 
more similar to the signal mode analysis, which has a missing neutrino.
The only significant background for the normalization analysis is the \pmz\ decay followed by
external conversion of one of the $\pi^0$ photons. 
This background is determined by MC simulations to be ($0.558\pm0.005$)\% 
of 1250828 normalization events after all cuts.


The acceptance ratio of \keteem\ to \pmzd\ depends on the efficiency ratios
of an electron and a charged pion, since the signal mode has one more 
electron and one less pion compared to the normalization mode.
Therefore, 
the efficiencies of an electron and a pion by the particle identification cuts are 
measured from data by tagging electrons and pions in \pmzd\ and \pmzgg\ decays, respectively.
The differences of the efficiencies between data and MC lead to a correction applied to the 
MC signal acceptance, which becomes $f_{e/\pi}=0.9955$.


After applying background subtractions and efficiency corrections, 
the decay width ratio of \keteem\ to \pmzd\ is
\begin{eqnarray}\label{eqn:R_ke3ee_pmzd}
&&\mathcal{R}({ke3ee}/{\rm +-0_{D}}) \nonumber\\
	&&\hspace{5mm}\equiv
	\frac{\Gamma(K_{e3ee}; 
	M_{e^+e^-} > 5 {\rm MeV}/c^2, E_{e^+e^-}^* > 30 {\rm MeV})}
	{\Gamma(K_L \to \pi^+\pi^-\pi^0_D)} \nonumber \\
	&&\hspace{5mm}= [8.54\pm0.07(stat)\pm0.13(syst)]\times 10^{-3},
\end{eqnarray}
where $E_{e^+e^-}^*$ is the energy of \eepair\ system in the kaon rest frame.
The acceptance of the signal events generated above the cut-off values is (0.8986$\pm$0.0025)\%,
and the acceptance of the normalization mode is (0.4947$\pm$0.0009)\%.
%
\begin{table}
\caption{\label{tbl:systematic}Systematic uncertainties in the 
ratio of decay widths, $\mathcal{R}({ke3ee}/{+-0_{D}})$, 
see Eq. \ref{eqn:R_ke3ee_pmzd} in the text.}
\begin{ruledtabular}
\begin{tabular}{ld}
Source of & \mbox{Uncertainty on} \\
uncertainty & \hspace*{-5mm}\mathcal{R}({ke3ee}/{+-0_{D}})(\%)\hspace*{-5mm} \\  
\hline
Unobserved photon & \\
\hspace{4mm}in normalization analysis & \pm 1.03\\
Vertex $\chi^2$ cut & \pm 0.7\\
Radiative corrections & \pm 0.51\\
Corrections for the efficiency difference  
& \pm 0.46 \\
$E_K$ distribution & \pm 0.35 \\
Cut-off on the $M_{e^+e^-}$ & - 0.18 \\
Background estimations & \pm 0.05 \\
MC statistics & \pm 0.32 \\
\hline
Total systematic uncertainties
 & \pm 1.5 \\
\end{tabular}
\end{ruledtabular}
\end{table}
%
Table \ref{tbl:systematic} lists the systematic errors in $\mathcal{R}({ke3ee}/{+-0_{D}})$.
The largest systematic error is
the uncertainty in the number of the \pmzd\ decays.
The number of \pmzd\ decays measured using the photon (full reconstruction measurement)
is (0.88$\pm$0.51)\% smaller than the analysis ignoring the photon.
With this value and the systematic error in the full reconstruction measurement of \pmzd,
we assign a 1.03\% systematic error on $\mathcal{R}({ke3ee}/{+-0_{D}})$.
The second largest systematic error is based on a slight data-MC discrepancy in the
distribution of the vertex $\chi^2$, which indicates the quality of the four track vertex.
The next largest systematic error is uncertainty in the treatment of radiative corrections. 
With inner bremsstrahlung photons generated in the MC by the {\sc photos} program,
the signal acceptance decreases by 3.6\%.
To confirm this acceptance loss, \ketee$\gamma$ (\keteem$_{\gamma}$) events 
are identified and compared between data and MC.
We assign a systematic uncertainty from the error in the \keteem$_{\gamma}$ measurement,
although the number of  \keteem$_{\gamma}$ evnents, 935, is consistent with MC prediction.
The probability to miss the $\pi$ track due to hadronic interactions in the TRD 
is determined by the {\sc geant} program \cite{Brun_GEANT:1994}.   
The correction applied to the MC signal acceptance is $1.0328\pm0.0045$.
We also estimate the uncertainties due to the \eop\ and TRD requirements.
The total error in our estimate of the efficiency difference is $\pm 0.46$\%.


The \keteem\ branching fraction with statistical and systematic uncertainty 
using $B$(\pmz)$=(12.56\pm0.05)\%$ 
and $B$(\pizeeg)$=(1.198\pm0.032)\%$ \cite{PDG2006} is
%
\begin{eqnarray} 
	&&B(K_{e3ee}; 
	M_{e^+e^-} > 5 {\rm MeV}/c^2, E_{e^+e^-}^* > 30 {\rm MeV}) \nonumber \\
	&&=  [1.285\pm0.010(stat)\pm0.040(syst)]\times10^{-5}.
\end{eqnarray}
%
The systematic error is much larger than that of $\mathcal{R}({ke3ee}/{+-0_{D}})$, 
due to the 2.7\% error on $B$(\pizeeg).


In the rest of this letter,
we compare our results against the ChPT[\nlopf] description of the \kpi\ current.
Using the \kesan\ branching fraction,
$B$(\kesan)$=(40.53\pm0.15)\%$ \cite{PDG2006}, we find
\begin{eqnarray}
	\mathcal{R}_{Ke3ee} &\equiv& 
	\frac{\Gamma(K_{e3ee}; 
	M_{e^+e^-} > 5 {\rm MeV}/c^2)}
	{\Gamma(K_{e3})}\nonumber \\
	&=&  [4.57\pm0.04(stat)\pm0.14(syst)]\times10^{-5}.
\end{eqnarray}
The leading order ChPT and ChPT[\nlopf] predictions for 
$\mathcal{R}_{ke3ee}$ are $(4.06\pm0.11)\times 10^{-5}$,
and $(4.29\pm0.11)\times 10^{-5}$ respectively.
Although the experimental determination of $\mathcal{R}_{Ke3ee}$ include all radiative effects,
neither the theoretical estimates of the numerator and denominator include these effects.
To account for the lack of radiative treatment in the $\mathcal{R}_{Ke3ee}$ predictions, 
we assign an error of twice the value of $\delta_{Ke3}$.  The variable $\delta_{Ke3}$ is  
$0.013 \pm 0.003$, and parameterizes the increase in decay width of the \kesan\ mode due to 
radiative corrections \cite{Alexopoulos:2004sw}.   The ChPT[\nlopf] prediction is 
consistent with the measurement at the 1.6 $\sigma$ level.


As the \kpi\ form factor is parameterized by the square of the four momentum transfer to the 
leptons  $t\equiv (p_K - p_{\pi})^2$, higher order calculations of ChPT are sensitive to $t$.
However, the \keteem\ decay has a two-fold ambiguity in $t$ due to the missing neutrino.
To avoid this problem, we use the transverse momentum transfer as defined in \cite{Alexopoulos:2004sy},
\begin{equation}
 t_{\perp} = M_K^2 + M_{\pi}^2 - 2 M_K \sqrt{p_{\perp,\pi}^2 + M_{\pi}^2},
\end{equation}
where $M_{\pi}$ is the charged pion mass and $p_{\perp,\pi}$ is the transverse pion momentum.
\par
\begin{figure}[tb]									
\begin{center}
\includegraphics[width=4.2cm]{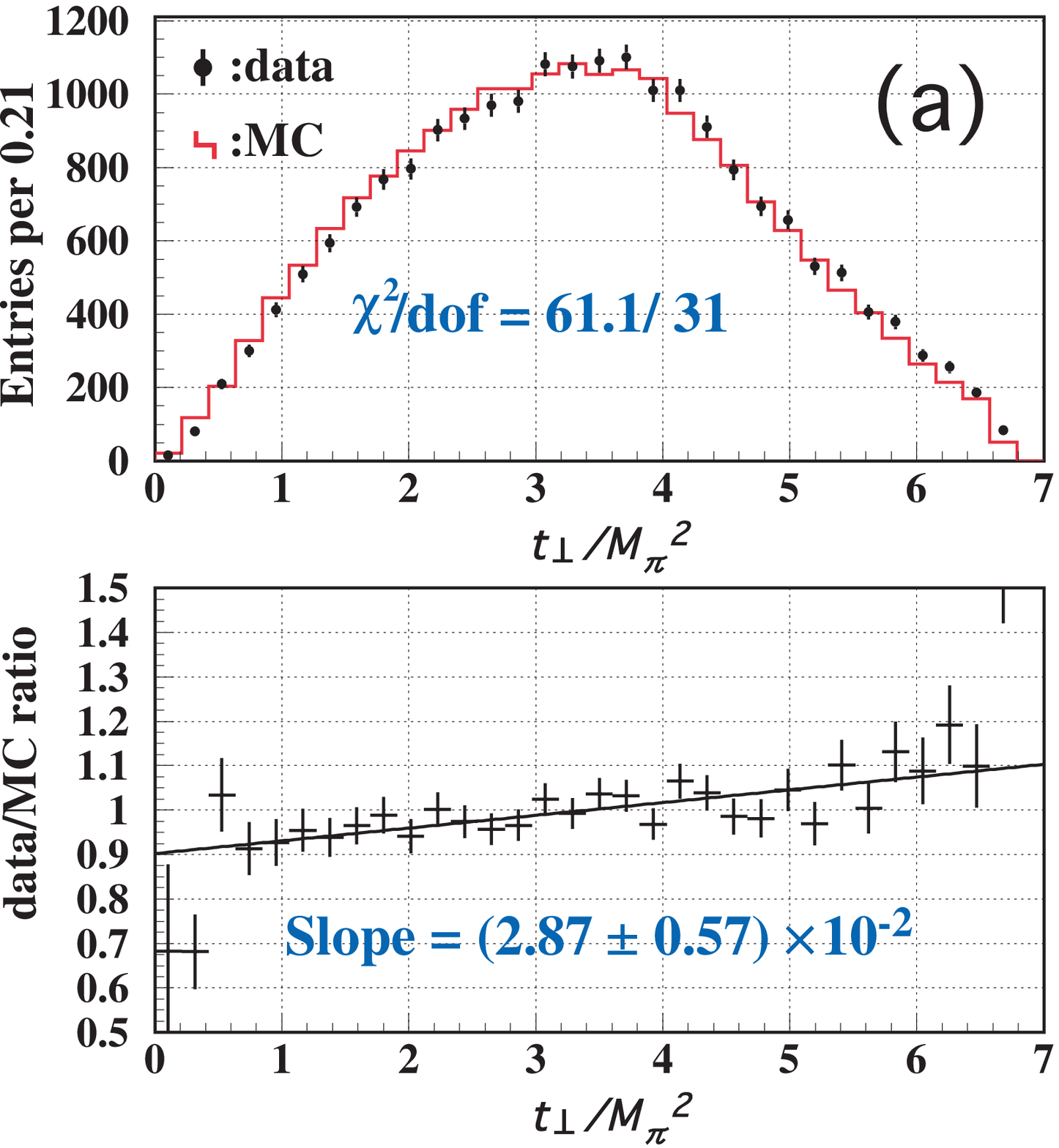}
\includegraphics[width=4.2cm]{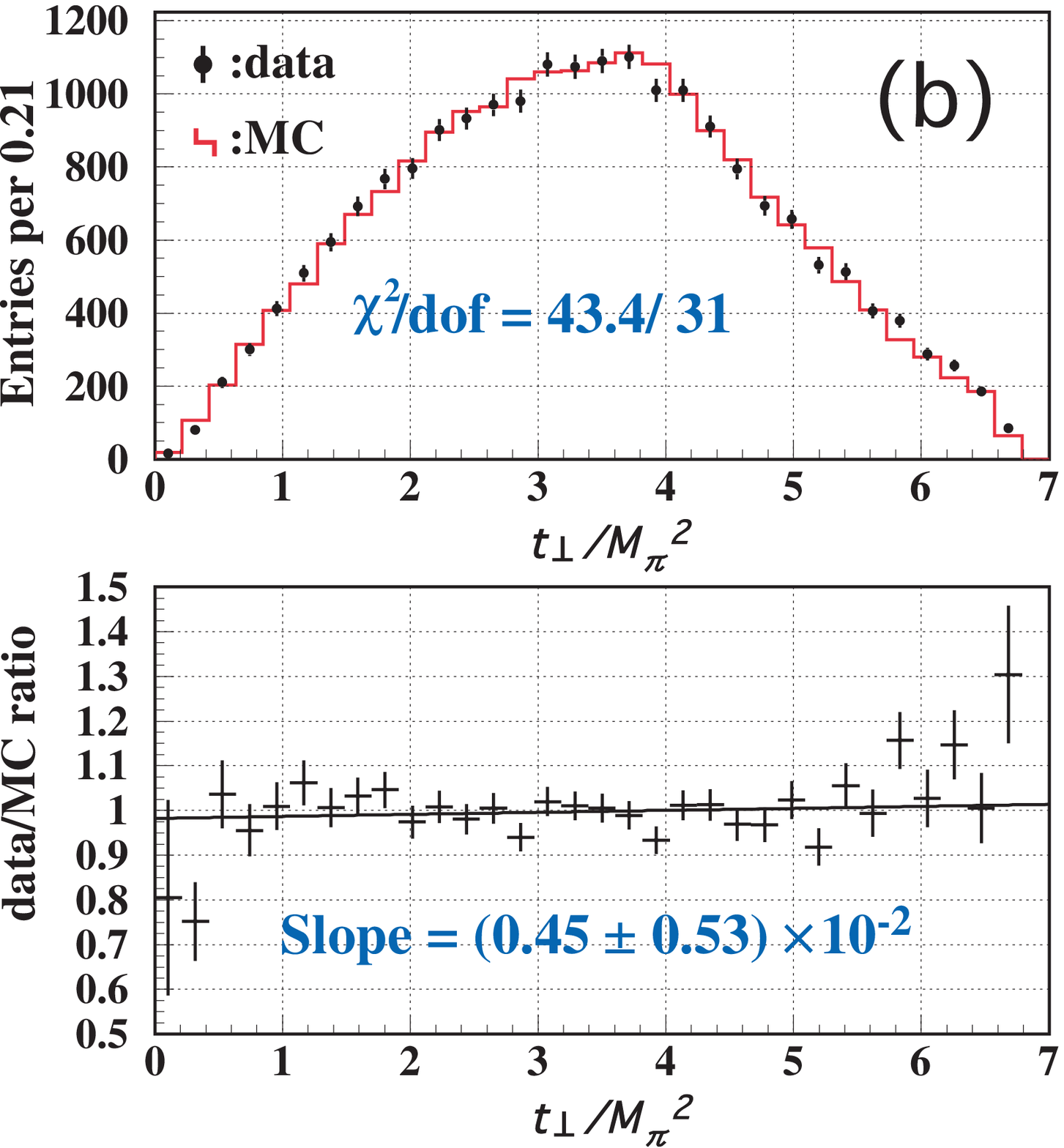}
\caption{\label{fig:t_pi_pape}
{Comparisons of the $t_{\perp}/M_{\pi}^2$ distributions for data (dots) and MC (histogram),
(a) with MC-LO and  (b) with MC-\nlopf. 
The data-to-MC ratios at the bottom are fit to a straight line.}}
\end{center} 
\end{figure}
\begin{figure}[htb]							
\begin{center}
\includegraphics[width=8.6cm]{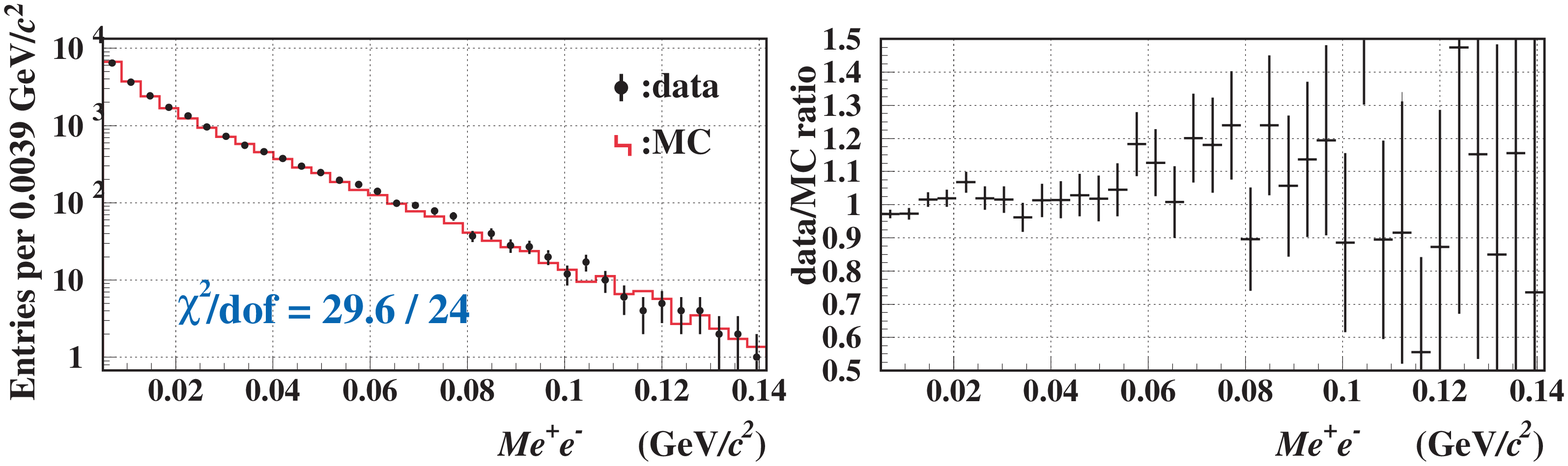} 
\caption{\label{fig:mee}
{ Comparison of the \mee\ distribution for data (dots), and MC (histogram) with \nlopf\ correction.
}}
\end{center} 
\end{figure}

Figure~\ref{fig:t_pi_pape} shows that the data $t_{\perp}/M^2$ distribution agrees with 
the \nlopf\ calculation, but not with the leading order ChPT calculation.
Figure~\ref{fig:mee} shows the invariant mass of the \eepair, 
illustrating that ChPT[\nlopf] models well the \keteem\ dynamics. 
The $M_{\pi eee}, M_{eee}$, and $M_{\pi e}$ distributions are also well modeled 
with the ChPT[\nlopf] prediction.

In summary, we find good agreement between our measurements and the \nlopf ChPT calculation, 
while the leading-order ChPT calculation is disfavored.  
Finally, we note that Figure \ref{fig:mee} is expected to receive 
contributions from both IB and SD amplitudes, with the IB amplitudes being dominant.  
The separation between IB and SD amplitudes has not been 
performed in the context of ChPT.  Additional theoretical work is needed to extract the SD 
contribution. 


We gratefully acknowledge the support and effort of the Fermilab
staff and the technical staffs of the participating institutions for
their vital contributions.  
We also thank K.~Tsuji and T.~Sato for useful discussions and calculating the
amplitude of the \keteem\ decay.
This work was supported in part by the U.S.
Department of Energy, The National Science Foundation, The Ministry of
Education and Science of Japan,
Fundacao de Amparo a Pesquisa do Estado de Sao Paulo-FAPESP,
Conselho Nacional de Desenvolvimento Cientifico e Tecnologico-CNPq and
CAPES-Ministerio Educao.


\newpage 

\newpage 

\end{document}